# The Structural Origin of Hydration Repulsive Force

Qiang Sun*

Key Laboratory of Orogenic Belts and Crustal Evolution, Ministry of Education, The School of

Earth and Space Sciences, Peking University

**ABSTRACT**

In our recent works, based on the structural studies on water and interfacial water (topmost water layer at the solute/water interface), hydration free energy is derived and utilized to investigate the physical origin of hydrophobic interactions. In this study, it is extended to investigate the structural origin of hydration repulsive force. As a solute is embedded into water, it mainly affects the structure of interfacial water, which is dependent on the geometric shape of solute. Therefore, hydrophobic interactions may be related to the surface roughness of solute. According to this study, hydration repulsive force can reasonably be ascribed to the effects of surface roughness of solutes on hydrophobic interactions. Additionally, hydration repulsive force can only be expected as the size of surface roughness being less than Rc (critical radius), which is in correspondence with the initial solvation process as discussed in our recent work. Additionally, this can be demonstrated by potential of mean force (PMF) calculated using molecular dynamics simulations.

* Corresponding author
  E-mail: QiangSun@pku.edu.cn

# 1 INTRODUCTION

Hydration repulsive forces exhibit for small separations and results in the surfaces to repel each other in water in the nanometer range. It was reported by LeNeveu et al. [1] measurement of the strong repulsion between neutral lipid bilayer membranes in distilled water. In fact, it is ubiquitous and acts between self-assembled membranes and surfactant layers [2], colloids [3], clays and biomolecules such as DNA [4], and proteins [5]. Therefore, hydration forces play an important role in many areas of science and engineering.

Experimentally, the interaction forces between two solutes can be directly measured by surface force apparatus (SFA) or atomic force microscope (AFM). To date, many experimental measurements have been conducted [6-16]. This means that this short-range force has universal character in the short range, less than 1 nm or about 3-4 water molecules, independent of solution conditions, that is, electrolyte ion size, charge ($Na^+$, $Ca^{2+}$, $Al^{3+}$) and concentration ($10^{-6}$-$10^{-2}$ M), and pH (3.9-8.2) [6]. Another striking feature is the common exponential dependence of the force on distance with an apparent decay length of about 0.25 nm [6-8], roughly the diameter of a water molecule, a consequence of a layer-by-layer dehydration of the surfaces when pushed together. This common force characteristic for these very different systems suggests that the force may be closely related to water structuring.

In general, the interaction between colloidal particles is viewed through the DLVO theory, which was named after Derjaguin, Landau, Verwey, and Overbeek, who developed the theory. It describes the surface forces between colloidal particles as the sum of the effects of the London-van der Waals attraction and the electrostatic repulsion due to the overlap of the double layer of counterions. The DLVO theory provides a successful explanation of the long-range interaction forces observed in a great variety of systems. However, when two surfaces or particles approach closer than a few nanometres, the repulsive interactions (hydration force) between two solid surfaces in a liquid medium fail to be accounted for by DLVO theory. Therefore, many works have been carried out to investigate the origin of hydration force [17-33].

To rationalize the hydration repulsive force, different mechanisms have been proposed. As discussed by Langmuir [34], an effective surface repulsion was suggested to arise from the enforced release of water molecules that are strongly bound to the membrane surfaces. This means that it is thermodynamically most favorable when a finite amount of hydration water is accommodated between the membrane surfaces. Additionally, this force is ascribed to the overlap of water ordering profiles at two opposing surface, which is theoretically shown to produce an exponentially decaying repulsion and reasoned to explain the universality of hydration forces observed for different surfaces [35,36]. In addition, according to Israelachvili and Wennerstrom work [18], they proposed that additional, direct surface interactions must play an equally important role for small surface separations.

The strength of hydrogen bonding of water is stronger than that of van der Waal interactions, it plays an important role in the process of hydration force. As a solute is dissolved into water, it mainly affects the structure of interfacial water layer (topmost water molecular layer at the solute/water interface) [37,38]. This means that the short-range hydration repulsive force may be related to the structured layer of water molecules at the solute surface. Due to the perturbations of water molecules by the macromolecules or surfaces, this may lead to the appearance of repulsive force.

In our recent work, based on the structural studies on interfacial water and bulk water [39-42],

hydration free energy is derived, and is utilized to investigate the physical origin of hydrophobic effects. In this study, it is extended to investigate the mechanism of hydration repulsive force. It can be derived that hydration force can be ascribed to the effects of surface roughness on hydrophobic interactions. Additionally, this can be demonstrated by potential of mean force (PMF) calculations using molecular dynamics simulations.

## 2 MOLECULAR DYNAMICS

### 2.1 MOLECULAR DYNAMICS SIMULATIONS

In this study, the molecular dynamics simulations are carried out using the program NAMD [43]. To simulate the surface roughness, a carbon atom is added on the surface of the graphite plane. Molecular dynamics simulations are conducted on different systems, each containing a target solute (a graphite plane, a graphite added a carbon atom roughness, a carbon atom) and a test solute (a $C_{60}$ fullerene, a methane molecule). The simulations were performed in the isobaric-isothermal ensemble (NPT). The simulated temperature was kept at 300 K, employing moderately damped Langevin dynamics. The pressure was maintained at 1 atm using a Langevin piston. The simulated box was 40Å×40Å×60Å. Additionally, periodic boundary conditions were applied in the three directions of Cartesian spaces.

The empirical CHARMM force field [44] was used to describe inter-atomic interactions. The intermolecular three point potential (TIP3P) model [45] was employed to represent the water molecules. Non-bonded van der Waals interactions were smoothly switched to zero between 10 and 12 Å. The PME algorithm was utilized to account for long-range electrostatic interactions. The equations of motion were integrated with a time step of 2 fs.

## 2.2 ABF CALCULATIONS

In this work, potential of mean force (PMF) was calculated using NAMD with the adaptive biasing force (ABF) [46-52] extensions integrated in the Collective Variables module. In the framework of ABF, the free energy along a transition coordinate can be treated as a potential resulting from the average force acting along the coordinate. In terms of generalized reaction coordinate $\xi$, the derivative of PMF can be expressed as,

$$\frac{dA(\xi)}{d\xi} = \left\langle \frac{\partial v(x)}{\partial \xi} - \frac{1}{\beta}\frac{\partial \ln|J|}{\partial \xi} \right\rangle_\xi = -\left\langle F_\xi \right\rangle_\xi \qquad (1)$$

where $|J|$ is the determinant of the Jacobian for the transformation from Cartesian to generalized coordinates, $v(x)$ is the potential energy function, and $F_\xi$ is the instantaneous force.

In the ABF method, a biasing force opposing the actual force arising from system components is periodically applied to the reaction coordinate to generate what is effectively a random walk along the reaction coordinate. The force applied along the reaction coordinate $\xi$, to overcome the PMF barriers is defined by,

$$F^{ABF} = \nabla_X \tilde{A}(\xi_0) = -\left\langle F_{\xi_0} \right\rangle_{\xi_0} \qquad (2)$$

Here, $\tilde{A}$ denotes the current estimate of the free energy and $\langle F_\xi \rangle_\xi$, the current average of $F_\xi$. As sampling of the phase space proceeds, the estimate $\nabla_x \tilde{A}$ is progressively refined.

In a simulation, the instantaneous force is calculated by dividing the reaction coordinate into discrete bins so that the average force exerted in the bin k is given by,

$$F_\xi(N_{step}, k) = \frac{1}{N(N_{step}, k)} \sum_{i=1}^{N(N_{step},k)} F_i(t_i^k) \qquad (3)$$

where $N(N_{step},k)$ is the number of samples collected in the bin k after $N_{step}$ simulation steps. $F_i(t_i^k)$ is the computed force at iteration i, and $t_i^k$ is the time at which the i*th* sample was collected in the bin k. For a sufficiently large $N_{step}$, $\overline{F}_\xi(N_{step},k)$ approaches the correct average force in each bin. Then, the free-energy difference, $\Delta A_\xi$, between the end point states can be estimated simply by way of summing the force estimates in individual bins,

$$\Delta A_\xi = -\sum_{i=1}^{M} \overline{F}_\xi\left(N_{step},k\right)\delta\xi \qquad (4)$$

In the study, the target solutes are fixed, and test solutes are restrained to move to the target solutes along Z axis. In the ABF calculations, the pathways between the solutes are broken down into several consecutive 2.8-3.0 Å wide windows.

## 3 DISCUSSIONS

### 3.1 HYDROPHOBIC INTERACTIONS

As the solutes are dissolved into water, the thermodynamic functions undoubtedly contain solute-solute, solute-water and water-water interaction energies,

$$\Delta G = \Delta G_{Water-water} + \Delta G_{Solute-water} + \Delta G_{Solute-solute} \qquad (5)$$

The strength of hydrogen bonding in water is stronger than that of van der Waals interactions, it is reasonable to ignore the Gibbs energy between solutes. Therefore, water plays an important role in the process of hydrophobic interactions. Additionally, it is necessary to study the structure of water, and the effects of the dissolved solute on water structure.

The formation of hydrogen bonding leads to the electronic transfer between neighboring water molecules, which shortens the O···O distance, but weakens the O-H covalent bond. Therefore, OH vibrational frequency is sensitive to hydrogen bonding, and widely applied to investigate the structure of liquid water. According to our recent studies [37-39], when three-dimensional hydrogen-bonded networks appear, OH vibration is mainly dependent on local hydrogen bonding (hydrogen bonding within the first shell) of a water molecule, and the effects of hydrogen bondings beyond the first shell on OH vibrations are weak. Therefore, different OH vibrations can be assigned to OH vibrations engaged in various local hydrogen-bonded networks of a water molecule.

Many works have been carried out to investigate the structure of water, which can roughly be divided into the mixture models and continuum models [53]. Based on our recent studies [37-39], at ambient conditions, a water molecule interacts with neighboring water molecules (the first shell) through various local hydrogen bondings, such as DDAA (double donor-double acceptor), DDA (double donor-single acceptor), DAA (single donor-double acceptor), and DA (single donor-single acceptor). Additionally, the hydrogen bondings of water are influenced by temperature, pressure, dissolved salt, and confined environments, which will be rearranged to oppose the changes of external conditions.

As a hydrophobic solute is embedded into water, an interface appears between the solute and water. The OH vibration is mainly dependent on the local hydrogen-bonded networks of a water molecule, the solute mainly affects the structure of topmost water layer at the solute/water interface (interfacial water). Due to the effects of dissolved solute on water structure, it can be into interfacial water and bulk water. Based on our recent studies [40], due to the truncations of hydrogen bondings at the solute/water interface, no DDAA (tetrahedral) hydrogen bonding can be found in the interfacial water, which is closely related to the interfacial formation.

When a solute is dissolved into water, it mainly affects the structure of interfacial water molecules. After the ratio of interfacial water layer to volume is determined, this is utilized to determine the loss of DDAA hydrogen bonding, and applied to calculate the Gibbs energy of interfacial water,

$$\Delta G_{Solute/water\ interface} = \Delta G_{DDAA} \cdot R_{Interfacial\ water/volume} \cdot n_{HB} \qquad (6)$$

where $\Delta G_{DDAA}$ is the Gibbs free energy of DDAA (tetrahedral) hydrogen bonding, $R_{Interfacial\ water/volume}$ is the molecular number ratio of interfacial water to volume, $n_{HB}$ is the average number of tetrahedral hydrogen bonding per molecule. For DDAA hydrogen bonding, $n_{HB}$ equals to 2.

Thermodynamically, the process of inserting a hard sphere into water is equivalent to that of forming an empty spherical cavity in water. After the solute is regarded as an ideal sphere, the molecular number ratio of interfacial water layer to volume is calculated to be $4 \cdot r_{H2O}/R$, where $r_{H2O}$ is the average radius of a H$_2$O molecule, and R is the radius of solute. Therefore, the hydration free energy can be determined as,

$$\Delta G_{Hydration} = \Delta G_{Water} + \Delta G_{Solute/water\ interface} = \Delta G_{Water} + \frac{8 \cdot \Delta G_{DDAA} \cdot r_{H2O}}{R} \qquad (7)$$

where $\Delta G_{Water}$ is the Gibbs free energy of pure water.

In principle, the lower hydration free energy, the more thermodynamically stable. Because hydration free energy is the sum of Gibbs energy of water and interfacial water, the structural transition can be expected as,

$$\Delta G_{Water} = \Delta G_{Solute/water\ interface} \quad (R = R_c) \qquad (8)$$

where Rc is the critical radius. In reference to Rc, it can be divided into the initial solvation and hydrophobic solvation processes, respectively. The Gibbs free energy of interfacial water is inversely proportional to the radius of the solute, which is proportional to the ratio of surface area to volume of dissolved solute. Therefore, different dissolved behaviors of solutes can be expected in various solvation processes.

In the initial solvation process, the Gibbs free energy of bulk water is larger than that of interfacial water, $\Delta G_{Solute/water\ interface} < \Delta G_{Water}$ (both of them are negative). To be more thermodynamically stable, the solutes may be dispersed in water to maximize $|\Delta G_{Solute/water\ interface}|$. It seems that there exists "repulsive" force between solutes. In the process of hydrophobic solvation, the Gibbs free energy of bulk water is lower than that of interfacial water, $\Delta G_{Water} < \Delta G_{Solute/water\ interface}$. To maximize the strength of bulk water $|\Delta G_{Water}|$, this undoubtedly leads to the solutes tend to be accumulated in water in order to minimize $|\Delta G_{Solute/water\ interface}|$. It seems that there exists "attractive" force between the solutes.

The hydration free energy is the sum of Gibbs energy of bulk water and interfacial water, therefore hydrophobic interactions (repulsive or attractive force) are closely related to the interfacial water. In fact, the interfacial water layer (or the solute/water interface) is closely related to the geometric shape of solute. Therefore, hydrophobic interactions may be influenced by the shape of solute, especially as the dissolved solute being composed of various structural units with different size, larger or less than critical size (Rc).

As a solute is dissolved into water, the Gibbs free energy between solute and water ($\Delta G_{Solute/water\ interface}$) is inversely proportional to the solute size, $\propto 1/\vec{R}$, where $\vec{R}$ is radius vector. When two solutes are embedded into water, the Gibbs energy of interfacial water should be inversely proportional to the separation between the two solutes, $\propto 1/\vec{R}_{Solute-solute}$, where $\vec{R}_{Solute-solute}$ is the separation between the solutes. Therefore, hydrophobic interactions may be closely related to the geometric characteristics of solutes, such as the solute size, solute shape, and the relative orientation between the solutes.

In this study, the test solute is a sphere, the target solute is a plane. When they are embedded into water (Figure 1(a)), based on our recent studies on hydrophobic effects [41,42], the test solute undoubtedly moves to the center of the target plane to be perpendicular to the plane in order to minimize the ratio of surface area to volume. With decreasing the separation between them, the hydration free energy can be expressed as,

$$\Delta G_{Hydration\ free\ energy} = \Delta G_{Bulk\ water} + \sum_{i=1}^{n} \Delta G_{Test\ solute-Target\ solute} = \Delta G_{Bulk\ water} + \iint \Delta G_{Test\ solute-Plane} dxdy \qquad (9)$$

where $\Delta G_{Test\ solute\text{-}plane}$ is the Gibbs energy of interfacial water arising from the solutes, and is inversely proportional to the separation between the solutes, $\propto 1/\vec{R}_{Test\ solute\text{-}plane}$.

In fact, the roughness can be expected to exist at the surface of any substance. In general, a solute can reasonably be divided into the plane substrate and the surface roughness. In this study, to characterize the surface roughness, a roughness is added at the center of the plane (Figure 1(b)). Because the $\Delta G_{Interfacial\ water}$ is related to the separation between the solutes, it can reasonably be expressed as,

$$\Delta G_{Test\ solute-Target\ solute} = \Delta G_{Test\ solute-Surface\ roughness} + \Delta G_{Test\ solute-Plane} + \Delta\Delta G_{Other} \qquad (10)$$

Mathematically, as the test solute approaches the target solute (a plane and a roughness), the hydration free energy can be divided into the sum of hydrophobic interaction between the solute and the plane ($\Delta G_{Test\ solute\text{-}Plane}$), and hydrophobic interactions between the solute and the surface roughness ($\Delta G_{Test\ solute\text{-}Surface\ roughness}$). Additionally, $\Delta\Delta G_{Other}$ is related to the difference of interfacial water layer as $\Delta G_{Test\ solute\text{-}Target\ solute}$ is divided into the sum of $\Delta G_{Test\ solute\text{-}Surface\ roughness}$ and $\Delta G_{Test\ solute\text{-}Plane}$.

In this study, after the plane is regarded as a circle (Figure 1(a)(b)), the $\Delta G_{Test\ solute\text{-}Plane}$ can be determined as,

$$\Delta G_{Test\ solute-Plane} \propto 1/R_{Test\ solute-Plane} = \pi \cdot \left[\sqrt{w^2 + (d+h)^2} - (d+h)\right] \qquad (11)$$

where w is the size of the circle, d is the separation between the test solute and roughness, h is the height of surface roughness, and d+h is the separation between the test solute and plane substrate. From this, $\Delta G_{Test\ solute-Plane}$ is dependent on the separation between test solute and plane, and the size of plane. Additionally, based on our recent works on hydrophobic interaction [42], the hydrophobic interactions are dependent on the solute size. Regarding to $\Delta G_{Test\ solute-Plane}$, the hydrophobic interactions between plane substrate and test solute can reasonably be regarded as "attractive" force.

Additionally, the hydrophobic interactions between the test solute and surface roughness ($\Delta G_{Test\ solute-Roughness}$) can be expressed as,

$$\Delta G_{Test\ solute-Roughness} \propto 1/d \qquad (12)$$

where d is the separation between the test solute and the roughness. However, different from the $\Delta G_{Test\ solute-Plane}$, the hydrophobic interactions between the test solute and surface roughness ($\Delta G_{Test\ solute-Roughness}$) may be "attractive" or "repulsive" force. This is dependent on the separation (d), and is also related to the $\Delta G_{Test\ solute-Roughness}$.

As the test and target solutes are embedded into water, hydration free energy is the sum of the Gibbs energy of bulk water and interfacial water, and the latter is composed of the interactions between test solute and plane substrate, and test solute and surface roughness. According to our recent works [41], the critical radius (Rc) can be determined as,

$$G_{Test\ solute-Surface\ roughness} + \Delta G_{Test\ solute-Plane} = \Delta G_{Bulk\ water} \quad (d = R_c) \qquad (13)$$

Under this condition, Rc is correspondence with the separation between the test solute and roughness. From the above equation, the Rc is affected by the $\Delta G_{Test\ solute-Plane}$. With increasing $\Delta G_{Test\ solute-Plane}$ (negative), this leads to the increase of the critical radius (Rc). Based on the equation (11), this can be fulfilled by increasing h or decreasing w. Additionally, the limit of critical radius can be determined when the $\Delta G_{\text{Test solute-Plane}}$ becomes zero. In fact, this changes into the hydrophobic interactions between two solutes. In our recent work [42], as two same solutes are dissolved into water, the repulsive force between solutes can be expected as the radius of solute being less than 3.2 Å at 300 K and 0.1 MPa.

In combination with our recent study on hydrophobic interactions [41], the hydrophobic interactions between the test solute and roughness may be dependent on the separation between them. As the separation (d) being larger than Rc, the "attractive" force can be expected to exist between the test solute and the roughness. However, hydrophobic interactions become "repulsive" only the separation is less than Rc. Therefore, the "repulsive" force can be found at short separations. This also indicates that the "repulsive" force can only be expected as the size both test solute and roughness being less than Rc. In other words, this means that the force is related to the size of surface roughness.

From the above, hydration repulsive force may be ascribed to the effects of surface roughness on hydrophobic interactions. Hydrophobic interactions may be "attractive" or "repulsive" force, which is closely related to the separation between the solute and surface roughness. This can be applied to understand that hydration force can only be detected at small separation between solutes. Additionally, this also indicates that hydration force may be expected as the size of solute being less than Rc. In fact, based on our recent studies [41], this is correspondence with the initial solvation process.

## 3.2 MOLECULAR DYNAMICS SIMULATIONS

Based on the above, hydration repulsive force can be ascribed to the effects of surface roughness on the hydrophobic interactions. In fact, this can be demonstrated by the PMFs calculated using molecular dynamics simulations. In this, study, the test solute is a $C_{60}$ fullerene. To investigate the effects of surface roughness on hydrophobic interactions, various target solutes are used, such as a single-layer graphite square plane (Figure 2(a)(b)), a graphite plane added a carbon atom (Figure 2 (c)(d)), and a single carbon atom(Figure 2(e)(f)), respectively. Additionally, the test solute is constrained to move to the center of the target solute to be vertical to the plane.

To investigate the water-induced contributions in the association of dissolved solutes, the potential of mean force (PMFs) between the solutes in water and vacuum are calculated by adaptive biasing force (ABF) calculations, respectively. From these, the water-induced contributions to the PMF can be determined as,

$$\Delta G_{Wter-induced} = \Delta G_{Solute-solute\,in\,water} - \Delta G_{Solute-solute\,in\,vacuum} \qquad (14)$$

Therefore, it can be utilized to investigate the hydrophobic interactions as the test solute approaches the target solute.

In this study, NAMD simulations are conducted on hydrophobic interaction between a graphite plane (no surface roughness) and a fullerene ($C_{60}$) (Figure 2(a)(b)). To minimize the ratio of surface area to volume, the $C_{60}$ fullerene moves to the center of graphite plate to be vertical to the plane. The PMFs can be determined from ABF calculations (Figure 3(a)(b)). This is in agreement with other studies on PMF calculations on graphite plates, $C_{60}$ fullerenes, and CNTs in water [54-56]. The first minimum in the PMF curve is referred to as the contact minimum. The second minima at separation of 10.0 Å is called the solvent-separated PMF, which corresponds to one water molecular layer is found between the solutes. Additionally, another minimum at the separation of 13 Å is observed, which corresponds to double water molecular layers between them.

From the water-induced PMFs, only attractive force can be detected as the $C_{60}$ fullerene approaches the graphite plate.

To investigate the effects of surface roughness on hydrophobic interactions, a carbon atom is added on the center of graphite surface (Figure 2(c)(d)). Additionally, the PMFs can be determined by ABF calculations when the $C_{60}$ fullerene approaching the roughness of target solute (Figure 3(c)(d)). In comparison with the hydrophobic interaction between the graphite plane and $C_{60}$ fullerene, it becomes repulsive force as the separation being less than 11 Å. Therefore, this is undoubtedly ascribed to the surface roughness on the hydrophobic interactions.

In this work, to further demonstrate the effects of surface roughness on hydrophobic interactions, molecular dynamics simulations are also conducted to investigate the hydrophobic interactions in the association of the carbon atom and $C_{60}$ fullerene (Figure 2(e)(f)). From figure 3(e)(f), the stronger repulsive force can be found in the PMF during the association of the $C_{60}$ fullerene with the carbon atom. Therefore, in comparison with smooth surface, as a carbon atom roughness is added on the plane, this undoubtedly results in the appearance of repulsive force at the small separation between solutes.

From the calculated PMFs, the repulsive force can be expected as the separation between solutes being smaller than the Rc (critical radius). Therefore, regarding to the structural origin of repulsive force, this arises from the enforced release of interfacial water molecules that are strongly bound to the surfaces. This means that it is more thermodynamically stable when a finite amount of hydration water is accommodated between the surfaces. According to our recent studies on hydrophobic interactions [41], this is in agreement with the initial solvation process. Because the Gibbs free energy of interfacial water is more thermodynamically stable than bulk water, this leads to the repulsive force as the interfacial water between the solutes are removed.

The hydration repulsive force can be expected as the separation between the test solute and roughness being less than Rc, which is related to the size both test solute and surface roughness. In this work, to further demonstrate the origin of repulsive force, after $CH_4$ molecule is used as test solute, the PMFs are also determined by ABF calculations (Figure 4). It can be found that repulsive force can still be detected at the small separation between the methane molecule and target solutes. Because the size of methane is less than that of $C_{60}$ fullerene, the repulsive force can also be expected between them.

In this work, after the target solute is divided into the plane and surface roughness, the hydrophobic interaction ($\Delta G_{\text{Test solute-target solute}}$) between the test and target solutes are regarded as the sum of the hydrophobic interactions between the test solute and plane substrate ($\Delta G_{\text{Test solute-plane}}$), and the solute and surface roughness ($\Delta G_{\text{Test solute-roughness}}$). To demonstrate the reasonability, based on the calculated PMFs, the sum of $\Delta G_{\text{Test solute-plane}}$ and $\Delta G_{\text{Test solute-roughness}}$ is determined (Figure 5). From figure 5, the calculated sum is roughly same as the $\Delta G_{\text{Test solute-target solute}}$ except for the repulsive force at small separation. In comparison with $\Delta G_{\text{Test solute-target solute}}$, the stronger repulsive can be found in the sum of calculated PMFs. The repulsive force is related to the enforced release of interfacial water, this indicates that stronger repulsive force corresponds to more interfacial water molecules. Regarding to target solute, interfacial water cannot be found between plane substrate and surface roughness, therefore the repulsive force of $\Delta G_{\text{Test solute-target solute}}$ is weaker than the calculated sum.

From equation (12), the strength of hydrophobic interactions is inversely proportional to the separation between the solutes. In reference to Rc, the hydrophobic interaction is attractive force as the separation being larger than Rc, and becomes repulsive force as the separation is less than Rc. Therefore, it can be derived that the hydration repulsive force may be dependent on the relative orientation between the test solute and surface roughness.

In this study, the ABF calculations are also conducted to study the effects of the orientation between the test solute and surface roughness on the hydration repulsive force. As the roughness moves from the center to the edge of plane (Figure 6), this results in the decrease of the hydration repulsive force of the calculated PMFs (Figure 7). This is in agreement with the above discussion on the dependence of hydrophobic interactions on the relative orientation between test solute and surface roughness. In fact, as the test solute approaches the target solute, there exist the "attractive" force between the solute and plane, and the "repulsive" force between the solute and roughness. As the roughness moves from the center to the edge of plane, the test solute will approach the plane more closely to the plane. Of course, this leads to the increase of attractive force between the solute and plane. Therefore, the strong hydration repulsive force can be expected as the test solute approaching the surface roughness.

According to the above discussion, hydration repulsive force can reasonably be ascribed to the effects of surface roughness on hydrophobic interactions, and the repulsive force can be expected as the surface roughness being less than Rc. In combination with our recent studies [41,42] on hydrophobic interactions, the "repulsive" force should be in correspondence with initial solvation process. According to recent study on the hydrophobic effects, this is dominated by the DA hydrogen bondings of interfacial water. Therefore, hydration force is thermodynamically attributed to the entropic process.

## 4 APPLICATIONS

As solutes are embedded into water, due to hydrophobic interactions, there exists the "attractive" or "repulsive" force between them. Regarding to hydration repulsive force, it can be ascribed to the effects of surface roughness on the hydrophobic interactions. It is inversely proportional to the separation between the test solute and roughness. To observe the hydration repulsive force, the separation must be less than Rc. In other words, the repulsive force can be expected as the size of surface roughness being less than Rc. This may be applied to understand and manipulate the hydrophobic interaction related to surface roughness.

To date, many surface force apparatus (SFA) and atomic force microscope (AFM) works have been conducted to measure the hydration force. However, different authors give various results. From the above discussion on hydration force, it can be derived that this may be related to the experimental methods. The hydration force can be expected at short separation, and the size of surface and tip roughness must be less than Rc. Therefore, the repulsive force may be influenced by the surface, the tip, or a combination of the two. In fact, this has been demonstrated by recent experimental studies [11,12,25-27].

In fact, the roughness can be found on any surface. If the separation between the roughness being less than Rc, hydration repulsive force can be expected to exist between well solvated surfaces in water at nanometer separations, such as the short-range repulsive interactions between DNA double helices. This force balances the van der Waals attraction in the nanometer range. It ultimately prevents the collapse of biological matter and thereby provides macromolecular assemblies with the necessary lubrication for vital functioning, even in the congested cell environment. The hydration repulsion between biological membranes creates a major barrier against close contact and thereby suppresses uncontrolled membrane adhesion and fusion. Therefore, the hydration repulsion is vital for the structural organization of cells and organelles as well as for their functionality.

In addition, this may be utilized to understand the effects of surface roughness on the hydrophobicity (or hydrophilicity). It has been found that hydrophobic properties are enhanced by increasing surface roughness [57,58]. This is measured as the roughness factor, which is defined as the ratio of the actual area of a rough surface to the geometric projected area. In fact, the increase of roughness factor is fulfilled by decreasing the separation (w) between neighboring roughness, and increasing the height (h) of roughness. Of course, these may lead to the decrease of the roughness size (Figure 8). Therefore, the roughness factor is in correspondence with the size of surface roughness. Based on equation (13), with decreasing w and increasing h, this leads to the increase of $R_c$, or increases the capacity of surface hydrophobicity. This can be applied to understand the self-cleaning (ultrahydrophobic) surface [59,60]. In other words, self-cleaning surface is expected as the surface roughness being less $R_c$ to keep the surface to be dry. In fact, this becomes the Cassie's model.

In our recent work [41], hydrophobic effects originate from the structural competition between hydrogen bonding in bulk water and that in interfacial water. Therefore, water plays an important role in the process of hydrophobic interactions. It is well known that the strength of hydrogen bonding of water is affected by temperature, pressure, the addition of NaCl, and the polarity of dissolved solutes, such as the hydrophobic or hydrophilic molecules. Therefore, it can be expected that hydration repulsive force may be affected by them.

## 5 CONCLUSIONS

In our recent works, based on the structural studies on water and interfacial water (topmost water molecular layer at the solute/water interface), hydration free energy is derived and utilized to investigate the physical origin of hydrophobic interactions. In the work, this is extended to investigate the structural origin of hydration repulsive force. Additionally, potential of mean force (PMF) is also calculated using molecular dynamics simulations to study the mechanism of hydration force. From the work, the following conclusions can be derived:

(1) Hydration repulsive force can be ascribed to the effects of surface roughness on hydrophobic interactions. Additionally, hydration force may be affected by the geometric shape of solute, such as the plane substrate, the height and width of roughness.

(2) Hydration force is only expected between the solutes as the roughness size of solutes being less than Rc (critical radius). In combination with our recent studies, the hydration repulsive force is in correspondence with the initial solvation process.

(3) The strength of hydration force is inversely proportional to the separation between the test solute and surface roughness. Additionally, it is also influenced by the relative orientation between them.

**ACKNOWLEDGEMENTS**

This work is supported by the National Natural Science Foundation of China (Grant Nos. 41773050).

**Figure 1.** (a) To minimize the ratio of surface area to volume, the test solute moves to the center of the target solute to be vertical to the plane. (b) As a roughness is added on the plane, the hydrophobic interactions can be divided into the interactions between the test solute and the plane, and those between the test solute and the surface roughness.

**Figure 2.** Different systems used to investigate the effects of surface roughness on hydrophobic interactions. Both initial and final configurations are shown. (a)(b) Target solute is the graphite plane. (c)(d) The roughness of a carbon atom is added on the plane. (e)(f) The target solute is a carbon atom. Water molecules are not shown for clarity.

**Figure 3.** (a) The PMFs in water and vacuum between the $C_{60}$ fullerene and graphite plane. (b) The water-induced PMF between the solutes. (c) The PMFs in water and vacuum as the $C_{60}$ fullerene moving to the plane with a carbon atom. (d) The corresponding water-induced PMF between the solutes. (e) The PMFs in water and vacuum as a $C_{60}$ fullerene approaching a carbon atom. (f) The corresponding water-induced PMF between the solutes.

**Figure 4.** The PMFs in water and vacuum (a) and the water-induced PMF (b) as a $CH_4$ molecule moving to the plane with a carbon atom roughness.

**Figure 5.** The sum of water-induced PMFs of $C_{60}$-plane and $C_{60}$-a carbon atom simulated systems. The water-induced PMF between $C_{60}$ fullerene and the plane with a carbon atom is also shown.

**Figure 6.** The dependence of PMFs on the orientation between the test solute and the roughness. Both initial and final configurations are shown. (a)(b) The carbon roughness is at the center of the plane. (c)(d) The roughness lies at the middle point between the center and edge of the plane. (e)(f) The roughness is at the edge of the plane. Water molecules are not shown for clarity.

**Figure 7.** (a) The PMFs in water and vacuum as the $C_{60}$ moving to the target solute where the roughness located at the center of the plane. (b) The water-induced PMF between the solutes. (c) The PMFs in water and vacuum as the $C_{60}$ moving to the target solute where the roughness located at the middle point between the center and edge of the plane. (d) The water-induced PMF between the solutes. (e) The PMFs in water and vacuum as the $C_{60}$ fullerene moving to the target solute where the roughness located at the edge of the plane. (f) The water-induced PMF between the solutes.

**Figure 8.** The dependence of surface hydrophobicity on the roughness.

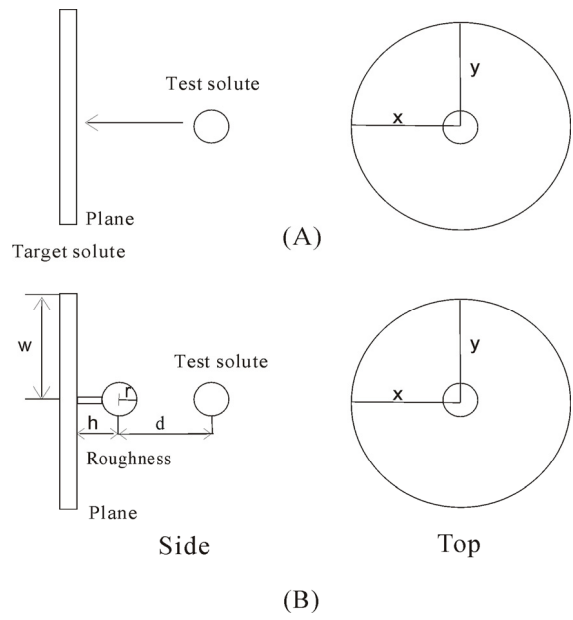

**Figure 1.**

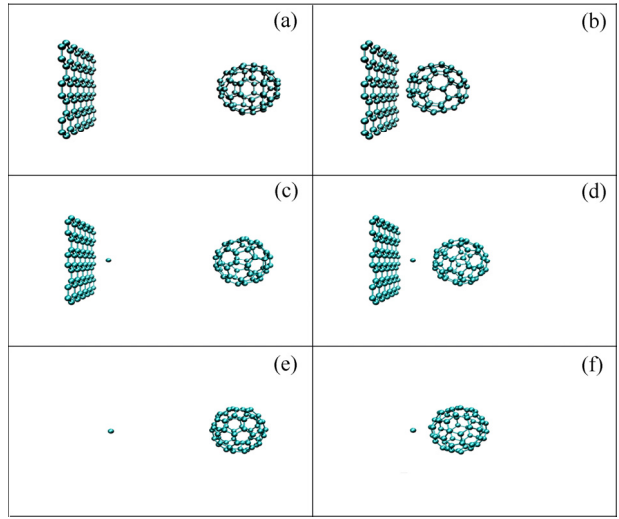

**Figure 2.**

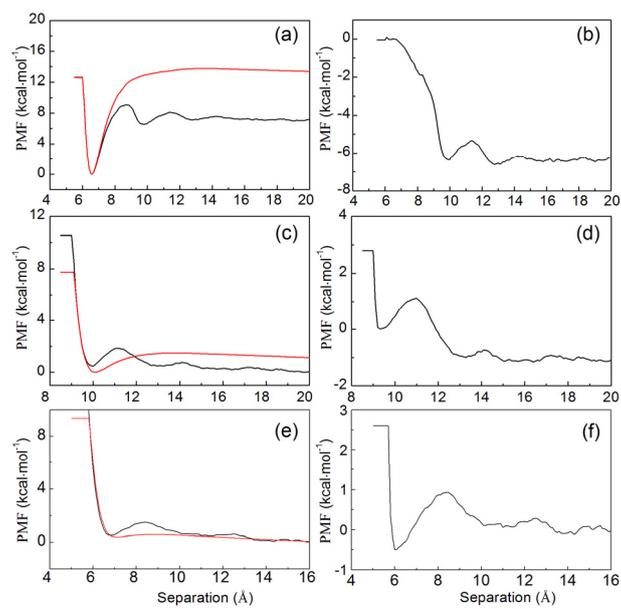

**Figure 3.**

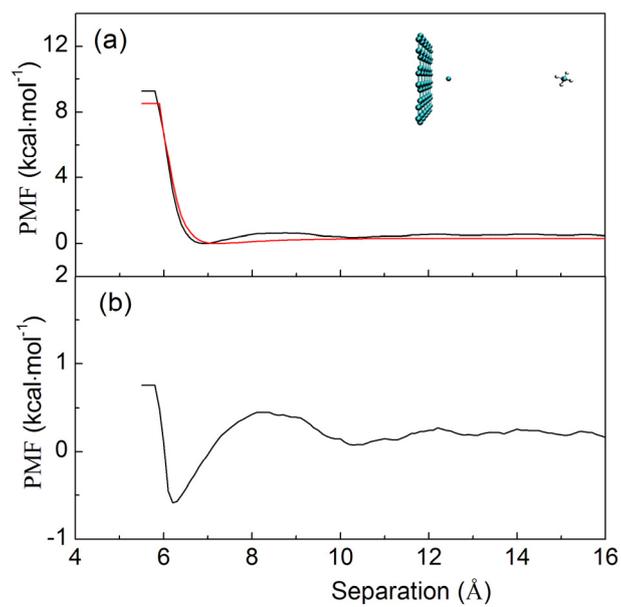

**Figure 4.**

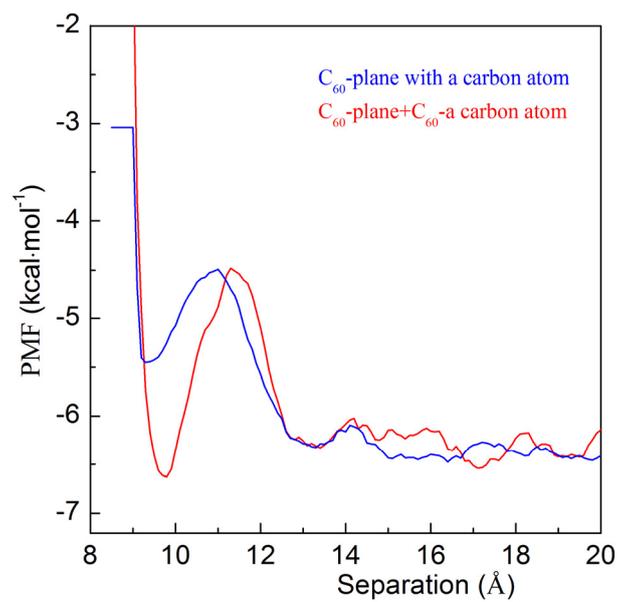

**Figure 5.**

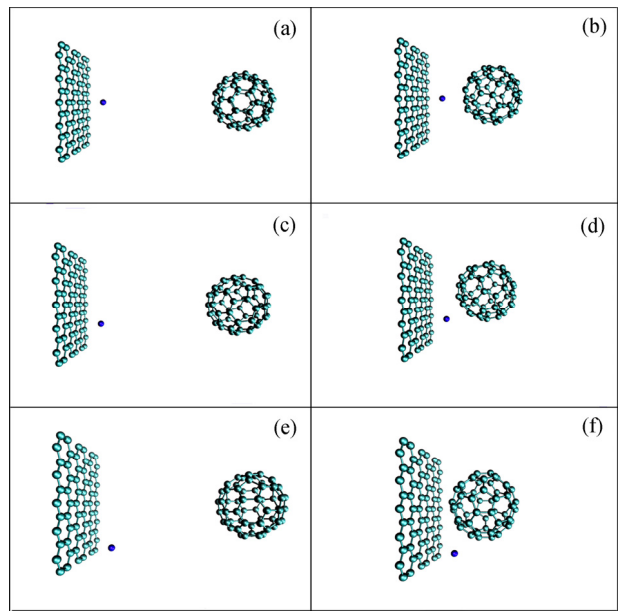

**Figure 6.**

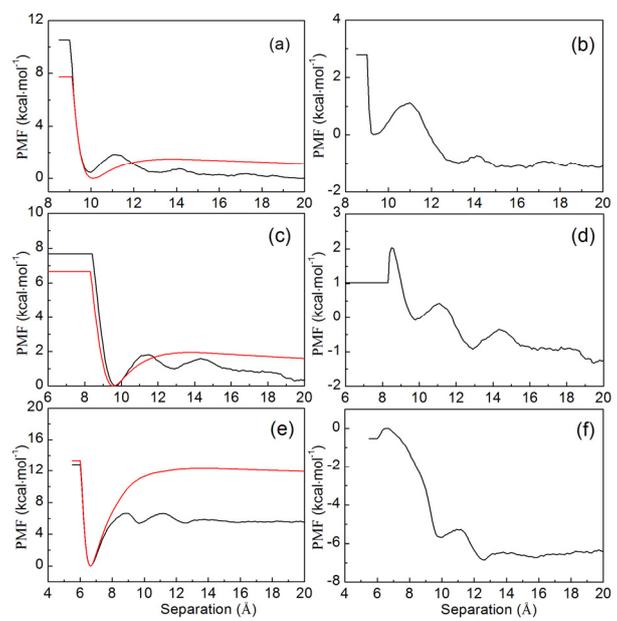

**Figure 7.**

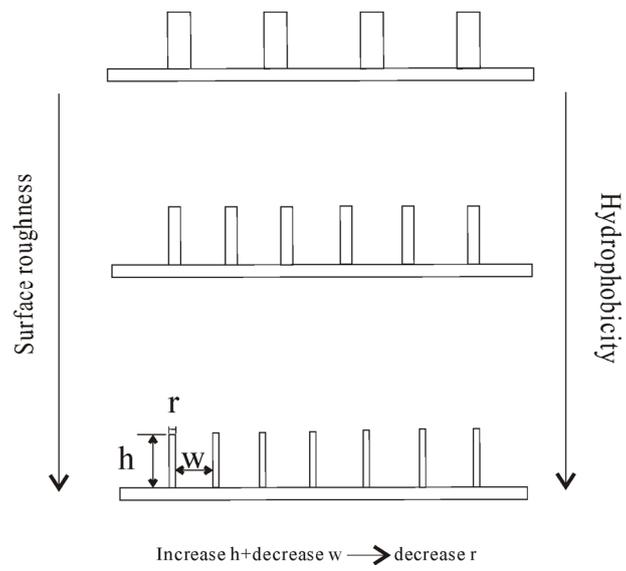

**Figure 8.**